\documentclass[aps,twocolumn]{revtex4}

\usepackage[english]{babel}
\usepackage[dvips]{graphicx}
\usepackage{color}
\usepackage{amssymb,amsfonts,amsmath}

\begin{document}



\title{Universality of citation distributions:
towards an objective measure of scientific impact.}

\author{Filippo Radicchi}
\affiliation
{Complex Networks Lagrange Laboratory (CNLL), ISI Foundation, Torino, Italy}

\author{Santo Fortunato}
\affiliation
{Complex Networks Lagrange Laboratory (CNLL), ISI Foundation, Torino, Italy}

\author{Claudio Castellano\thanks{To whom correspondence should be addressed.
E-mail: claudio.castellano@roma1.infn.it}}
\affiliation{SMC, INFM-CNR, and Dipartimento di Fisica,
``Sapienza'' Universit\`a di Roma, Piazzale A. Moro 2, 00185 Roma, Italy}





\begin{abstract}
We study the distributions of citations received by a single publication
within several disciplines, spanning broad areas of science.
We show that the probability that an article is cited $c$ times has large
variations between different disciplines, but all distributions are
rescaled on a universal curve
when the relative indicator $c_f=c/c_0$ is considered, where $c_0$ is the
average number of citations per article for the discipline. 
In addition we show that the same universal behavior occurs when
citation distributions of articles published in the same field, but in
different years, are compared.
These findings provide a strong validation of $c_f$ as an unbiased
indicator for citation performance across disciplines and years.
Based on this indicator, we introduce a generalization of the h-index
suitable for comparing scientists working in different fields.
\end{abstract}


\maketitle
\section{Introduction}

Citation analysis is a bibliometric tool that is becoming
increasingly popular to evaluate the performance of different actors in the
academic and scientific arena, ranging from individual
scholars~\cite{hirsch05, egghe06, hirsch07}, to journals, departments,
universities~\cite{evidence07} and national institutions~\cite{kinney07} up
to whole countries~\cite{king04}.
The outcome of such analysis often plays a crucial role to decide
which grants are awarded, how applicants for a position are ranked,
even the fate of scientific institutions.
It is then crucial that citation analysis is carried out in the most
precise and unbiased way.

Citation analysis has a very long history and many potential problems
have been identified~\cite{brooks86,egghe90,adler08}, the most critical being
that often a citation does not -- nor it is intended to -- reflect
the scientific merit of the cited work (in terms of quality or relevance).
Additional sources of bias are, to mention just a few,
self-citations, implicit citations, the increase in the total number
of citations with time or the correlation between the number of authors
of an article and the number of citations it receives~\cite{bornmann08}.

In this work we consider one of the most relevant factors that may hamper
a fair evaluation of scientific performance: field variation.
Publications in certain disciplines are typically cited much more or much less
than in others. This may happen for several reasons,
including uneven number of cited papers per article in different fields
or unbalanced cross-discipline citations~\cite{althouse08}.
A paradigmatic example is provided by mathematics: the highest 2006
impact factor (IF)~\cite{garfield79} for journals in this category
(Journal of the American Mathematical Society) is 2.55, whereas
this figure is ten times larger or even more in other disciplines
(for example, New England Journal of Medicine has 2006 IF 51.30,
Cell has IF 29.19, Nature and Science have IF 26.68 and 30.03, respectively).

The existence of this bias is well-known~\cite{garfield79,egghe90,bornmann08}
and it is widely recognized that comparing bare citation numbers
is inappropriate. Many methods have been proposed to
alleviate this
problem~\cite{schubert86,schubert96,vinkler96,vinkler03,iglesias07}. 
They are based on the general idea of normalizing
citation numbers with respect to some properly chosen reference standard.
The choice of a suitable reference standard, that can be a journal,
all journals in a discipline or a more complicated set~\cite{schubert96}
is a delicate issue~\cite{zitt05}.
Many possibilities exist also in the detailed implementation of the
standardization procedure. Some methods are based on ranking articles
(scientists, research groups) within one field and comparing
relative positions across disciplines.
In many other cases {\em relative indicators} are defined, i.e. ratios
between the bare number of citations $c$ and some average measure
of the citation frequency in the reference standard.
A simple example is the Relative Citation Rate of a group of
articles~\cite{schubert86},
defined as the total number of citations they received, divided
by the weighted sum of impact factors of the journals
where the articles were published. 

The use of relative indicators is widespread, but
empirical studies~\cite{naranan71,seglen92,redner98} 
have shown that distributions of article citations are very skewed,
even within single disciplines. 
One may wonder then whether it is appropriate to normalize by the
average citation number, that gives only very limited characterization
of the whole distribution.
We address this issue in this article.

The problem of field variation affects the evaluation of performance
at many possible levels of detail: publications, individual scientists,
research groups, institutions.
Here we consider the simplest possible level,
the evaluation of citation performance of single publications.
When considering individuals or research groups, additional sources
of bias (and of arbitrariness) exist, that we do not tackle here.
As reference standard for an article, we consider the set of all papers
published in journals that are classified in the same Journal
of Citation Report scientific category
of the journal where the publication appears (see details in 
Sec.~\ref{methods}).
We take as normalizing quantity for citations of articles
belonging to a given scientific field the average number $c_0$
of citations received by all articles in that discipline published
in the same year.
We perform an empirical analysis of the distribution of citations for
publications in various disciplines and we show that the large
variability in the number of bare citations $c$ is fully accounted for
when $c_f=c/c_0$ is considered. The distribution of this relative performance
index is the same for all fields.
No matter whether, for instance, Developmental Biology, Nuclear Physics or
Aerospace Engineering are considered, the chance of having a particular
value of $c_f$ is the same.
Moreover, we show that $c_f$ allows to properly take into account
the differences, within a single discipline, between articles published in
different years.
This provides a strong validation of the use of $c_f$
as an unbiased relative indicator of scientific impact for
comparison across fields and years.

\section{Variability of citation statistics in different disciplines}

\begin{figure}
\begin{center}
\resizebox{0.9\columnwidth}{!}{\includegraphics{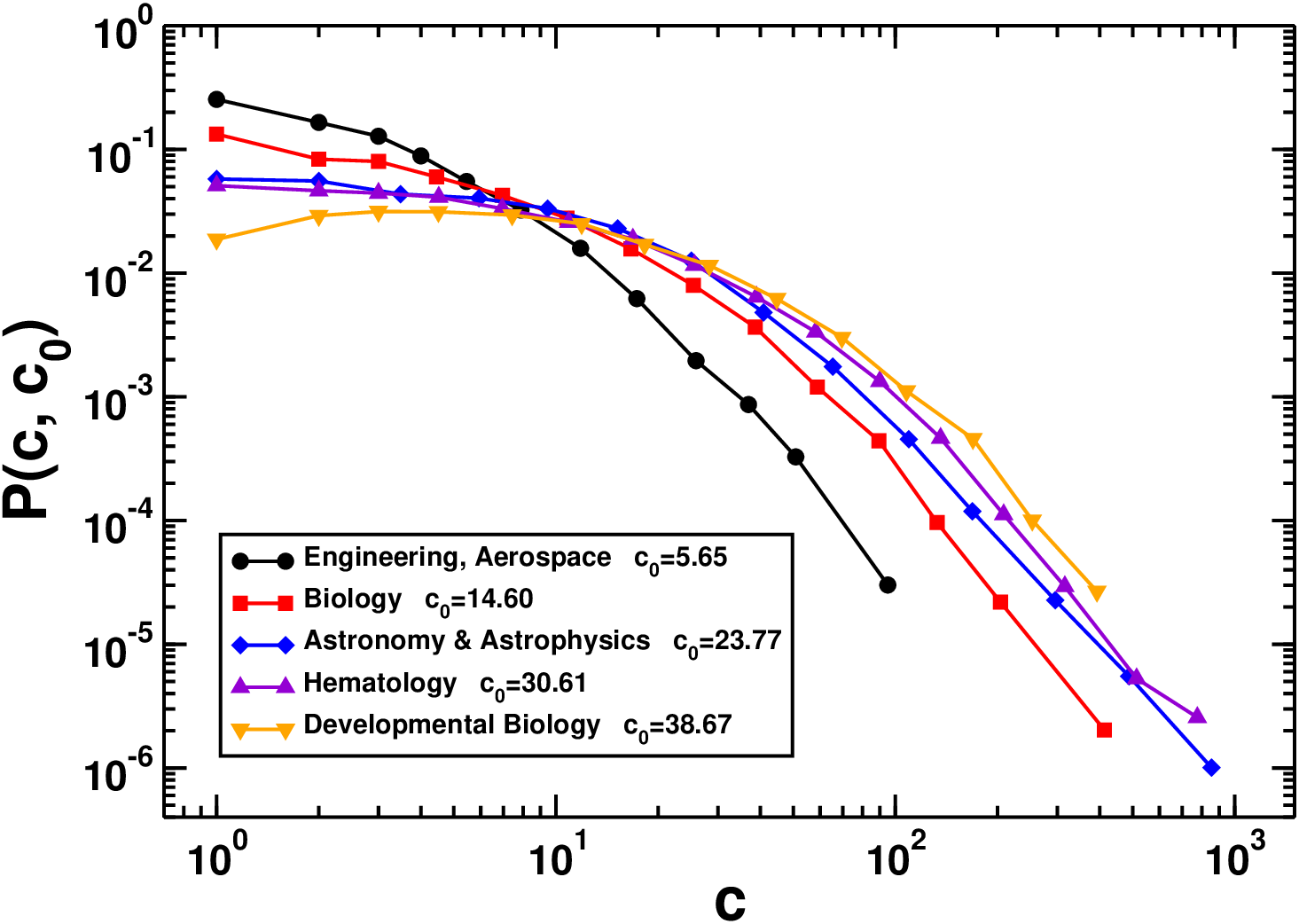}}
\end{center}
\caption{Normalized histogram of the number of articles $P(c,c_0)$ published
in 1999 and having received $c$ citations.
We plot $P(c,c_0)$ for several scientific disciplines with different average
number $c_0$ of citations per article.}
\label{pv}
\end{figure}

First of all we show explicitly that the distribution of the number
of articles published in some year and cited a certain number of times
strongly depends on the discipline considered.
In Fig.~\ref{pv} we plot the normalized distributions of citations
to articles that appeared in 1999 in all journals belonging to several different
disciplines according to the Journal of Citation Reports
classification.

From this figure it is apparent that the chance of a publication
to be cited strongly depends on the category the article belongs to. 
For example a publication with 100 citations is approximately 50 times
more common in ``Developmental Biology'' than in ``Engineering, Aerospace''.
This has obvious implications in the evaluation of outstanding scientific
achievements: the simple count of the number of citations is patently
misleading to assess whether an article in Developmental Biology is more
successful than one in Aerospace Engineering.

\section{Distribution of the relative indicator $c_f$}

\begin{figure}
\begin{center}
\resizebox{0.85\columnwidth}{!}{\includegraphics{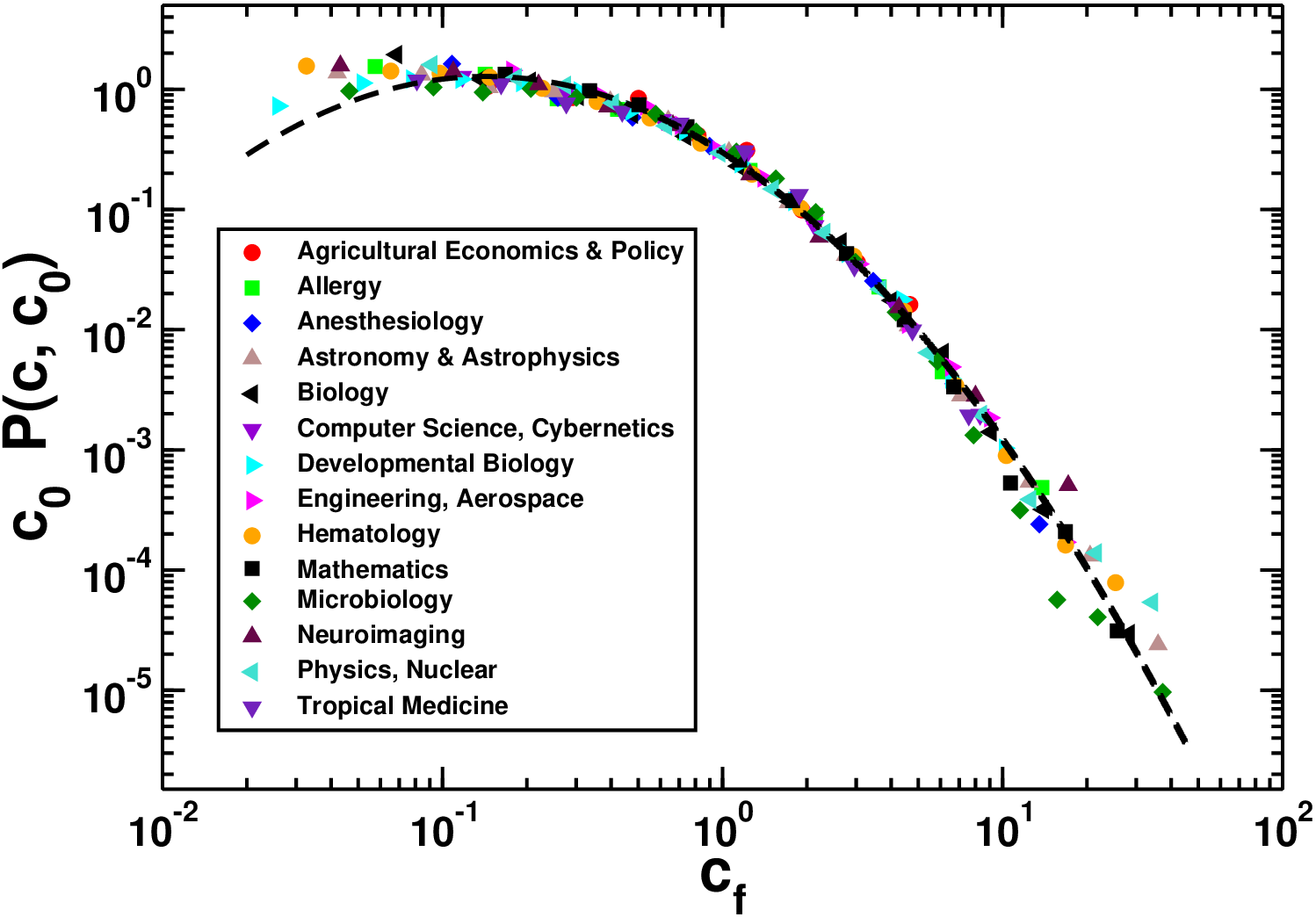}}
\end{center}
\caption{Rescaled probability distribution $c_0 P(c,c_0)$ of the relative
indicator $c_f=c/c_0$, showing that the universal scaling holds
for all scientific disciplines considered (see table~\ref{table}).
The dashed line  is a lognormal fit with $\sigma^2=1.3$.}
\label{pv0}
\end{figure}

A first step toward properly taking into account field variations is
to recognize that the differences in the bare citation distributions are 
essentially not due to specific discipline-dependent factors, but are
instead related to the pattern of citations in the field, as measured
by the average number of citations per article $c_0$.
It is natural then to try to factor out the bias induced by the difference
in the value of $c_0$ by considering a relative indicator,
i.e. measuring the success of a publication by the ratio $c_f=c/c_0$ between the
number of citations received and the average number of citations 
received by articles published in its field in the same year.
Fig.~\ref{pv0} shows that this procedure leads to a very good collapse
of all curves for different values of $c_0$ onto a single shape.
The distribution of the relative indicator $c_f$ seems then universal
for all categories considered and resembles a lognormal distribution.
%
In order to make these observations more quantitative, we have fitted
each curve in Fig.~\ref{pv0} for $c_f \ge 0.1$ with a lognormal curve
\begin{equation}
F(c_f)=\frac{1}{\sigma c_f \sqrt{2\pi}}e^{-{(\log(c_f)-\mu)^2}/2\sigma^2},
\label{lognormal}
\end{equation}
where the relation $\sigma^2=-2\mu$, due to the fact that the expected value
of the variable $c_f$ is 1, reduces the number of fitting parameters to one.
All fitted values of $\sigma^2$, reported in Table~\ref{table}, are compatible
within two standard deviations, except for one (Anesthesiology) that is in
any case within three standard deviations of all the others.
Values of $\chi^2$ per degree of freedom, also reported in Table~\ref{table},
indicate that the fit is good. 
%
\begin{table*}
\begin{center}
\begin{tabular}{c||r || c | c | c | c | c | c }
Index & Subject Category & year& $N_p$  & $c_0$ &  $c_{max}$  & $\sigma^2$ &  $\chi^2/d_f$
\\
\hline
1&Agricultural Economics \& Policy  & 1999 &    266&6.88&	42      &       1.0(1)  &       0.007
\\
\hline
2&Allergy      &   1999 &  1530	&	17.39	&		271     & 	1.4(2)  &      	0.012
\\
\hline
3&Anesthesiology & 1999 &   3472&	13.25	&	282	&	1.8(2)  &      	0.009 
\\ 
\hline
4&Astronomy \& Astrophysics   &  1999 &  7399 & 23.77&		1028	&	1.1(1)  &      	0.003
\\
\hline
5&Biology   &  1999 &        3400&	14.6	&		413	&	1.3(1)  &     	0.004 
\\
\hline
6&Computer Science, Cybernetics &  1999 &  704 & 8.49&	100	&	1.3(1)	&	0.004
\\
\hline
7&Developmental Biology      & 1999 &  2982& 38.67	&	520     &	1.3(3)	&	0.002
\\
\hline
8&Engineering, Aerospace    & 1999 & 1070&	5.65	&	95	&	1.4(1)	&	0.003
\\
\hline
9&Hematology     &  1990 &  4423&	41.05	&		1424	&	1.5(1)	&	0.002
\\
\hline
10&Hematology     &  1999 &  6920&	30.61	&		966	&   	1.3(1)	&	0.004
\\
\hline
11&Hematology     &  2004 & 8695&	15.66	&		1014	&	1.3(1)	&	0.003
\\
\hline
12&Mathematics & 1999 & 8440	&	5.97	&	191	&	1.3(4)	&	0.001
\\
\hline
13&Microbiology    &  1999 &  9761&	21.54	&	803	&	1.0(1)	&	0.005 
\\
\hline
14&Neuroimaging   &  1990 &    444&	25.26	&	518	&	1.1(1)	&	0.004
\\
\hline
15&Neuroimaging   &  1999 &   1073&	23.16	&	463	&	1.4(1)	&	0.003   
\\
\hline
16&Neuroimaging   &  2004 &    1395&	12.68	&	132	&	1.1(1)	&	0.005
\\
\hline
17&Physics, Nuclear     &  1990 &  3670 &  13.75&	387	&	1.4(1)	&	0.001
\\
\hline
18&Physics, Nuclear     &  1999 &   3965&10.92	&	434	&	1.4(4)	&	0.001
\\
\hline
19&Physics, Nuclear     &  2004 &  4164	&6.94	&	218	&	1.4(1)  &	0.001 
\\
\hline
20&Tropical Medicine  &  1999 &    1038	&12.35	&	126	&	1.1(1)	&	0.017
\\

\end{tabular}
\end{center}
\caption{List of all scientific disciplines considered in this
  article. For each category we report the total number of articles $N_p$,
  the average number of citations $c_0$,
  the maximum number of citations $c_{max}$,
  the value of the fitting parameter $\sigma^2$ in Eq.~(\ref{lognormal})
  and the corresponding $\chi^2$ per degree of freedom. Data refer to
  articles published in journals listed by Journal of
  Citation Reports under a specific subject category.}
\label{table}
\end{table*}
%
This allows to conclude that, rescaling the distribution of citations
for publications in a scientific discipline by their average number,
a universal curve is found, independent of the specific discipline.
Fitting a single curve for all categories, a lognormal distribution
with $\sigma^2=1.3$ is found, that is reported in Figure~\ref{pv0}.

Interestingly, a similar universality for the distribution of the
relative performance is found, in a totally different context, when the
number of votes received by candidates in proportional elections is
considered~\cite{fortunato07}. 
In that case, the scaling curve is also well-fitted by a lognormal with
parameter $\sigma^2 \approx 1.1$.
For universality in the dynamics of academic research activities see
also~\cite{plerou99}.

\begin{figure}
\begin{center}
\includegraphics[width=0.45\textwidth]{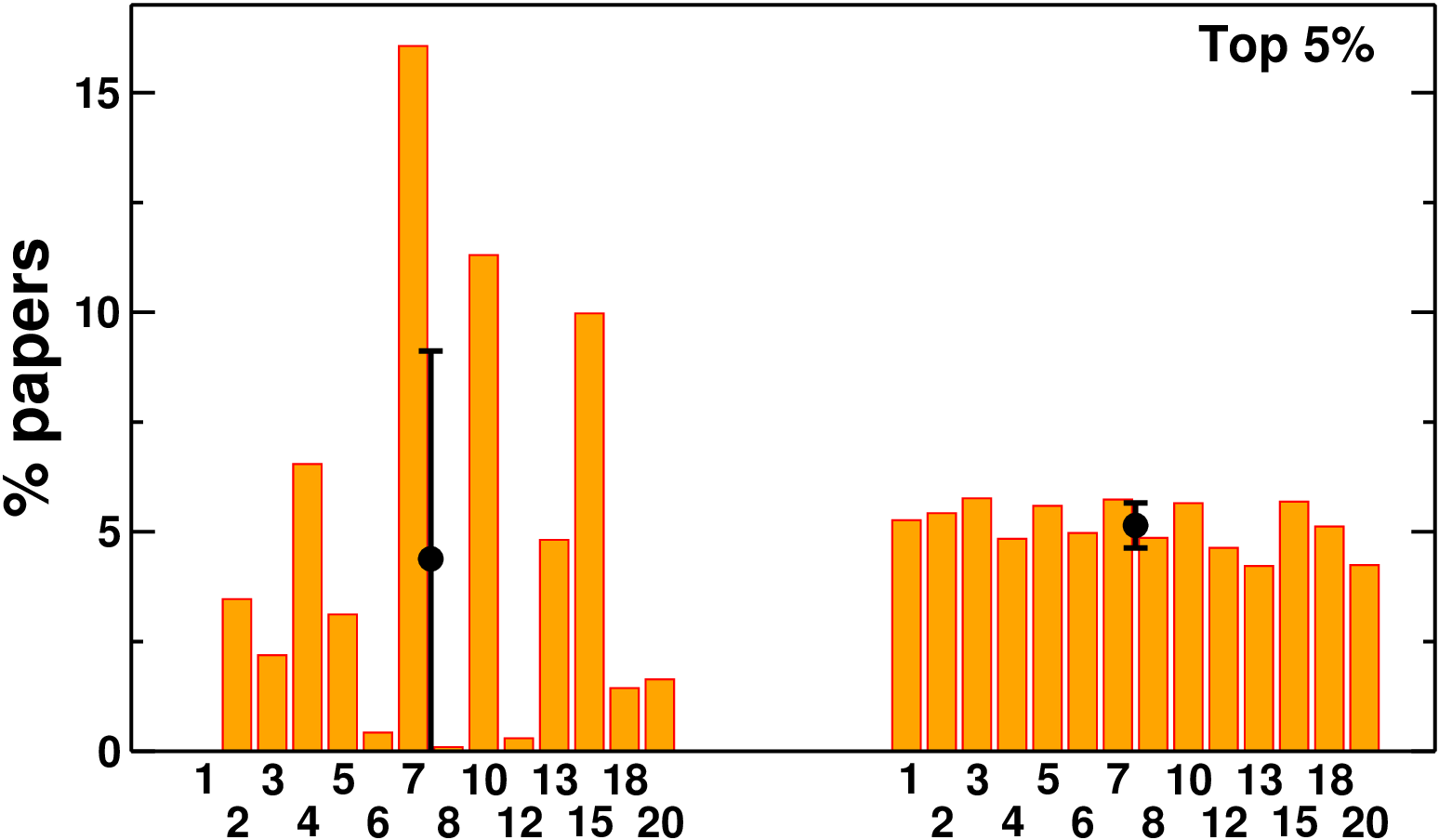}
\quad
\includegraphics[width=0.45\textwidth]{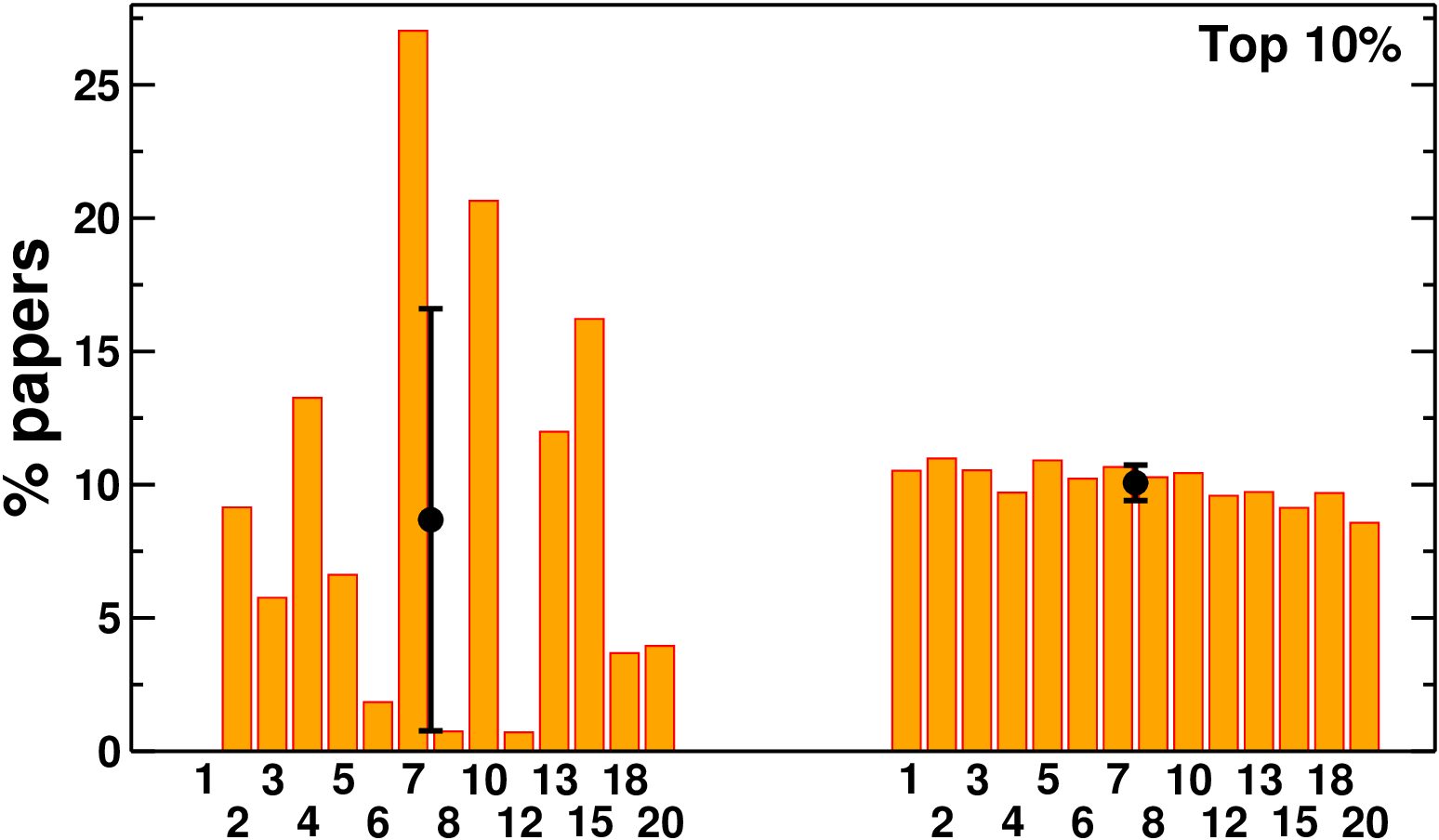}
\\
\vspace{0.4cm}
\includegraphics[width=0.45\textwidth]{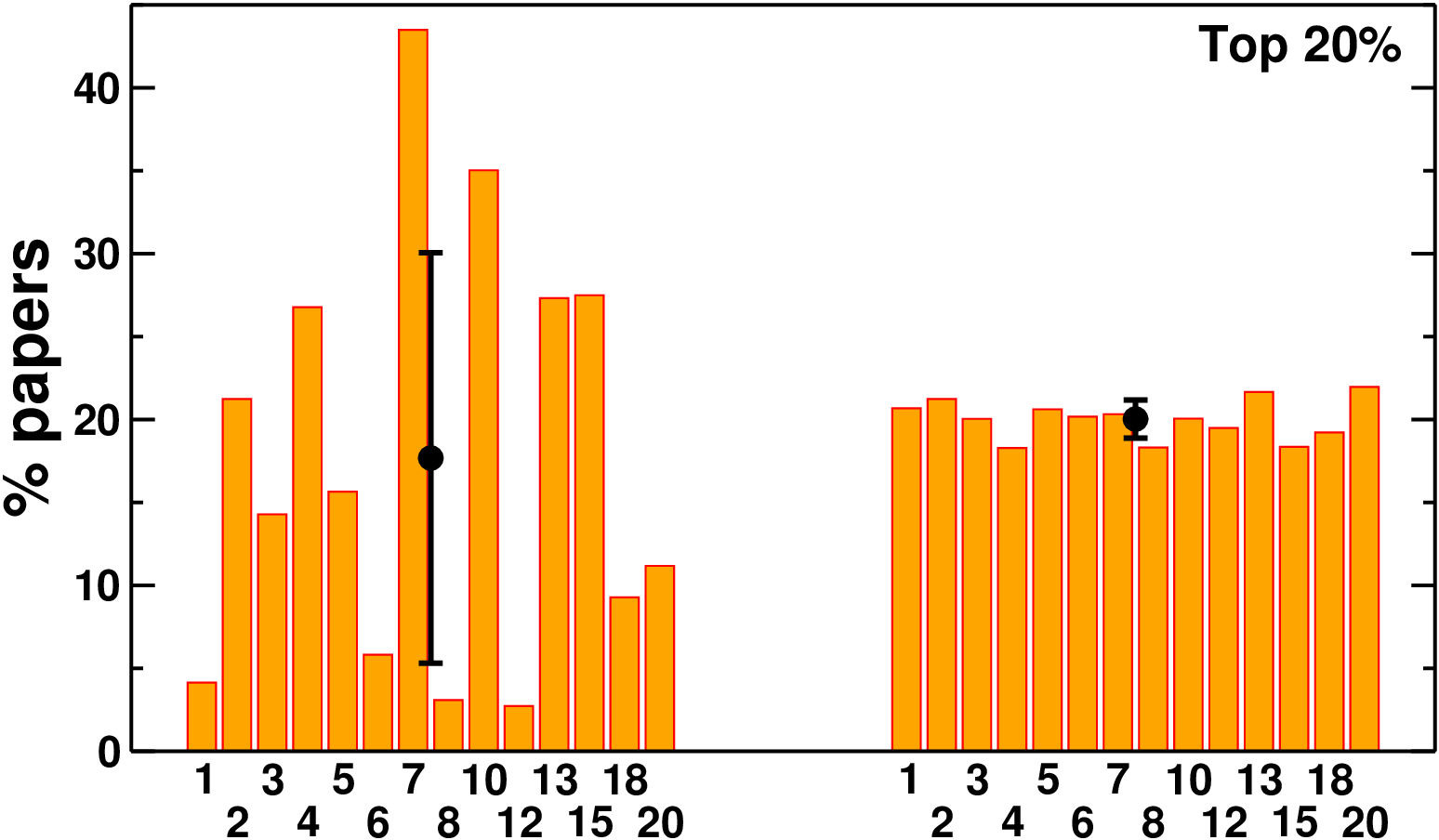}
\quad
\includegraphics[width=0.45\textwidth]{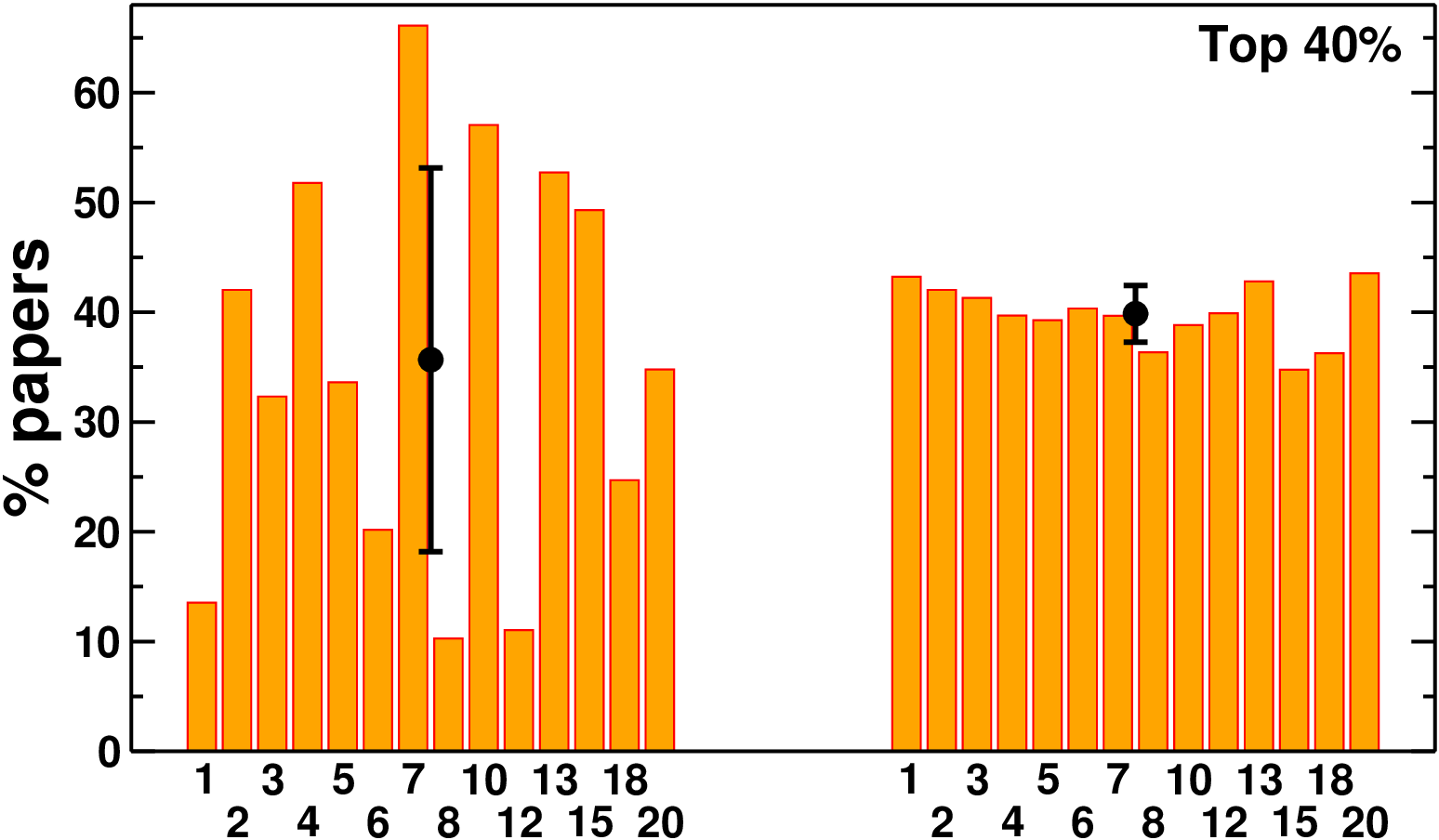}
\\
\vspace{1cm}
\end{center}
\caption{We rank all articles according to the bare number of citations
  $c$ and the relative indicator $c_f$. We then plot the percentage of
  articles of a particular discipline present in the top $z\%$ of the
  general ranking, for the rank based on the number of citations (
  histograms on the left in each panel) and based on the relative
  indicator $c_f$ (histograms on the right).
  Different values of $z$ (different panels) lead to very similar pattern of
  results. The average values and the standard deviations of the bin
  heights shown are also reported in Table~\ref{table2}. The numbers
  identify the disciplines as they are indicated in
  Table~\ref{table}.}
\label{fig6new}
\end{figure}

The universal scaling obtained provides a solid grounding for
comparison between articles in different fields. To make this even
more visually evident, we have ranked all articles belonging to a pool
of different disciplines (spanning broad areas of science) according
either to $c$ and to $c_f$.  We have then computed the percentage of
publications of each discipline that appear in the top $z\%$ of the
global rank. If the ranking is fair the percentage for each discipline
should be around $z\%$ with small fluctuations.  Fig.~\ref{fig6new}
clearly shows that when articles are ranked according to the
unnormalized number of citations $c$ there are wide variations among
disciplines. Such variations are dramatically reduced instead when the
relative indicator $c_f$ is used. This occurs for various choices of the
percentage $z$. More quantitatively, assuming that articles of the
various disciplines are scattered uniformly along the rank axis, one
would expect the average bin height in Fig.~\ref{fig6new} to be $z\%$
with a standard deviation
\begin{equation}
\sigma_z = \sqrt{ \frac{z
  \left(100-z\right)}{N_c} \sum_{i=1}^{N_c} \frac{1}{N_i}},
\end{equation}
where $N_c$ is the number of categories and $N_i$ the number of articles
in the $i$-th category.  When the ranking is performed according to
$c_f=c/c_0$ we find (Table~\ref{table2}) a very good agreement with
the hypothesis that the ranking is unbiased, while strong evidence
that the ranking is biased is found when $c$ is used. For example, for
$z=20\%$, $\sigma_z = 1.15\% $ for $c_f$-based ranking, while
$\sigma_z = 12.37 \% $ if $c$ is used, as opposed to the value
$\sigma_z = 1.09 \%$ in the hypothesis of unbiased ranking.
\begin{table}
\begin{center}
\begin{tabular}{c|c||c|c||c|c}
z       &    $\sigma_z(theor)$ & $z(c)$ & $\sigma_z(c)$ & $z(c_f)$ & $\sigma_z(c_f)$ \\
\hline
5       &     0.59   &  4.38 &  4.73 &  5.14 &      0.51    \\
10      &     0.81   &  8.69 &  7.92 & 10.07 &      0.67    \\
20      &     1.09   & 17.68 & 12.37 & 20.03 &      1.15    \\
40      &     1.33   & 35.67 & 17.48 & 39.86 &      2.58    \\

\end{tabular}
\end{center}
\caption{Average and standard deviation for the bin heights in Fig.~\ref{fig6new}. Comparison
between the values expected theoretically for unbiased ranking (first two columns), and those
obtained empirically when
articles are ranked according to $c$ (third and fourth columns) and according to $c_f$ (last two columns).}
\label{table2}
\end{table}
Figures~\ref{pv0} and~\ref{fig6new} allow to conclude that $c_f$ is an
unbiased indicator for comparing the scientific impact of publications
in different disciplines.
%
For the normalization of the relative indicator, we have considered
the average number $c_0$ of citations per article published in the same year
and in the same field. This is a very natural
choice, giving to the numerical value of $c_f$
the direct interpretation as relative citation performance of
the publication.
In the literature this quantity is also indicated as the 
``item oriented field normalized citation score''~\cite{lundberg07},
an analogue for a single publication of the popular
CWTS (Centre for Science and Technology Studies, Leiden)
field normalized citation score or ``crown indicator''~\cite{moed95}.
In agreement with the findings of Ref.~\cite{althouse08}
$c_0$ shows very little correlation with the overall size of the field,
as measured by the total number of articles.

\begin{figure}
\begin{center}
\resizebox{0.9\columnwidth}{!}{\includegraphics{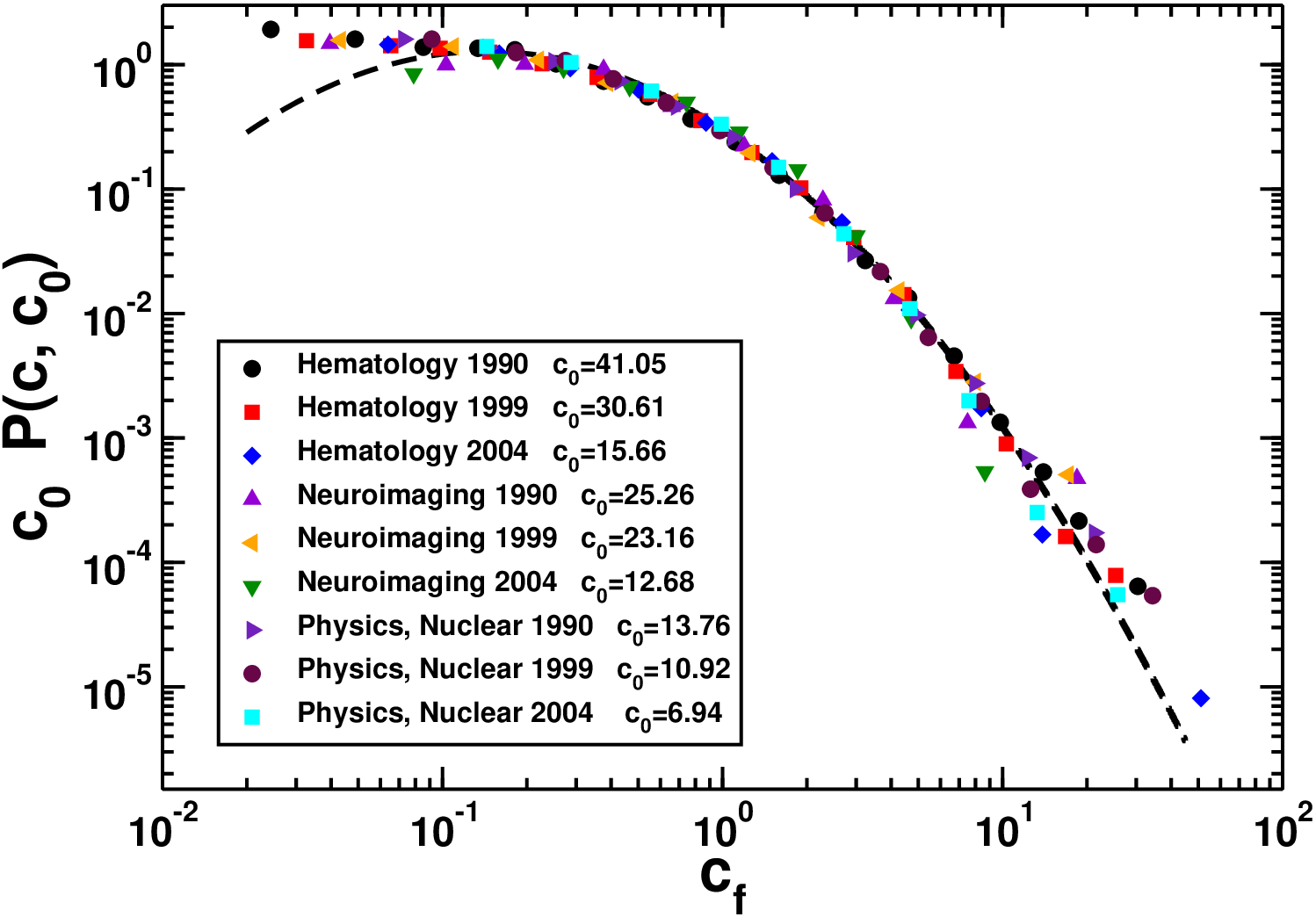}}
\end{center}
\caption{Rescaled probability distribution $c_0 P(c,c_0)$ of the relative
indicator
$c_f=c/c_0$ for three disciplines (``Hematology'',  ``Neuroimaging'', and
"Physics, Nuclear'') for articles published in
different years ($1990$, $1999$ and $2004$). In spite of
the natural variation of $c_0$ ($c_0$ grows as a function of the
elapsed time), the universal scaling observed over different
disciplines naturally holds also for articles published in different
periods of time.
The dashed line  is a lognormal fit with $\sigma^2=1.3$.}
\label{pv0years}
\end{figure}

The previous analysis compares distributions of citations to articles
published in a single year, 1999. It is known that different 
temporal patterns of citations exist, with some articles starting soon
to receive citations, while others (``sleeping beauties'') go unnoticed
for a long time, after which they are recognized as seminal and begin
to attract a large number of citations~\cite{vanraan04,redner05}.
Other differences exist between disciplines, with noticeable fluctuations
in the cited half-life indicator across fields.
It is then natural to wonder whether the universality of distributions
for articles published in the same year extends longitudinally in time
so that the relative indicator allows comparison of articles
published in different years.
For this reason, in Fig.~\ref{pv0years} we compare the plot of
$c_0 P(c,c_0)$
vs $c_f$ for publications in the same scientific discipline appeared
in three different years. The value of $c_0$ obviously grows as older
publications are considered, but the rescaled distribution remains
conspicuously the same.

\section{A generalized h-index}

Since its introduction in 2005, the h-index~\cite{hirsch05} has
enjoyed a spectacularly quick success~\cite{ball05}:
it is now a well established standard tool for the evaluation of the
scientific performance of scientists.
Its popularity is partly due to its simplicity: the h-index of an author
is $h$ if $h$ of his $N$ articles have at least $h$ citations each, and
the other $N - h$ articles have at most $h$ citations each.
Despite its success, as all other performance metrics the h-index has
some shortcomings, as already pointed out by Hirsch himself.
One of them is the difficulty to compare authors in different
disciplines.
\begin{figure}
\begin{center}
\resizebox{0.9\columnwidth}{!}{\includegraphics{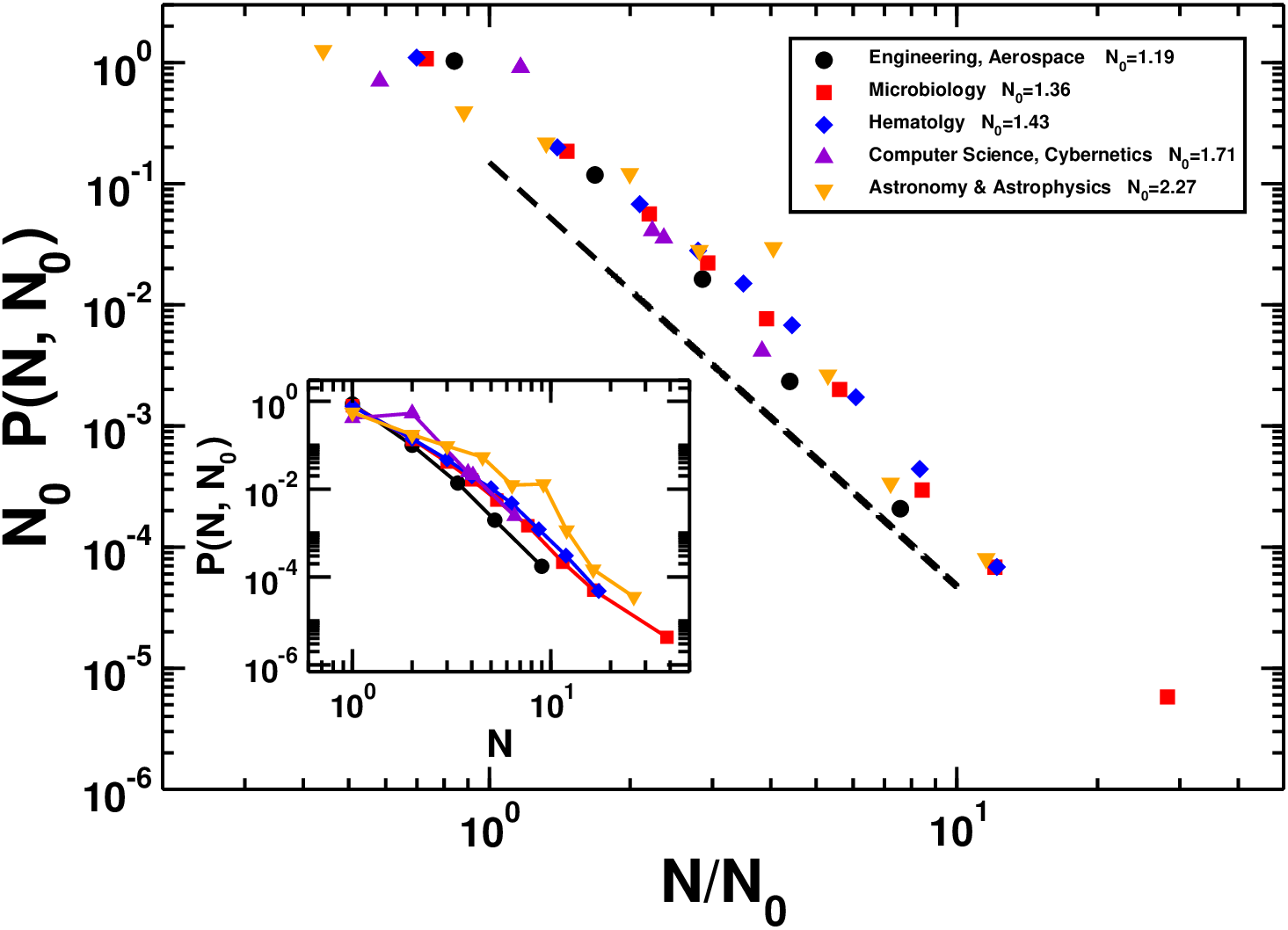}}
\end{center}
\caption{
Inset: distributions of the number of articles $N$ published by
an author during 1999 in several disciplines. 
Main: the same distributions rescaled by the average number $N_0$
of publications per author in 1999 in the different disciplines.
The dashed line is a power-law with exponent $-3.5$.}
\label{PN0}
\end{figure}
The identification of the relative indicator $c_f$ as the correct
metrics to compare articles in different disciplines naturally suggests
its use in a generalized version of the h-index taking properly
into account different citation patterns across disciplines.
However, just ranking articles according to $c_f$, instead of on the basis
of the bare citation number $c$, is not enough. A crucial ingredient
of the h-index is the number of articles published by an author.
As Fig.~\ref{PN0} shows, such a quantity also depends on the
discipline considered: in some disciplines the average number
of articles published by an author in a year is much larger than
in others. But also in this case this variability is
rescaled away if the number $N$ of publications in a year by an
author is divided by the average value in the discipline $N_0$.
Interestingly, the universal curve is fitted reasonably well over almost
two decades by a power-law behavior
$P(N,N_0) \approx \left(N/N_0\right)^{-\delta}$
with $\delta=3.5(5)$.

This universality allows one to define a generalized h-index, $h_f$,
that factors out also the additional bias due to different publication rates,
thus allowing comparisons among  scientists working in different fields.
To compute the index for an author, his/her articles are ordered according
to $c_f=c/c_0$ and this value is plotted versus the reduced rank $r/N_0$
with $r$ being the rank.
In analogy with the original definition by Hirsch, the generalized index
is then given by the last value of $r/N_0$ such that the
corresponding $c_f$ is larger than $r/N_0$.
For instance, if an author has published 6 articles with values of
$c_f$ equal to 4.1, 2.8, 2.2, 1.6, 0.8 and 0.4 respectively, 
and the value of $N_0$ in his discipline is 2.0, his
$h_f$-index is equal to 1.5. This because the third best article has
$r/N_0=1.5<2.2=c_f$, while the fourth has $r/N_0=2.0 > 1.6=c_f$.
We plan to present the results of the application of this generalized
index to practical cases in a forthcoming publication.

\section{Conclusions}

In this article we have presented strong empirical evidence that the widely
scattered distributions of citations for publications in different
scientific disciplines are rescaled upon the same universal curve
when the relative indicator $c_f$ is used. We have also seen that the
universal curve is remarkably stable over the years.
The analysis presented here justifies the use of relative
indicators to compare in a fair manner the impact of
articles across different disciplines and years.
This may have strong and unexpected implications.
For instance, Figure~\ref{pv0} leads to the counterintuitive conclusion 
that an article in Aerospace Engineering with only 20 citations
($c_f \approx 3.54$) is more successful than an article in
Developmental Biology with 100 citations
($c_f \approx 2.58$). We stress that this does not
imply that the article with larger $c_f$ is  necessarily
more ``important'' than the other. In an evaluation of importance,
other field-related factors may play a role: an article with an
outstanding value of $c_f$ in a very narrow specialistic field may be
less "important'' (for science in general or for the society)
than a publication with smaller $c_f$ in a highly competitive
discipline with potential implications in many areas.

Since we consider single publications, the smallest possible entities
whose scientific impact can be measured, our results must always be
taken into account when tackling other, more complicated tasks, like
the evaluation of performance of individuals or research groups.
For example, in situations where the simple count of the mean number
of citations per publication is deemed to be important, one should
compute the average of $c_f$ (and not of $c$) to evaluate impact
independently of the scientific discipline.
For what concerns the assessment of single authors' performance
we have defined a generalized h-index~\cite{hirsch05} that allows a fair
comparison across disciplines taking into account also the different
publication rates.

Our analysis deals with two of the main sources of bias affecting
comparisons of publication citations. It would be interesting to tackle,
along the same lines, other potential sources of bias, as
for example the number of authors, that is known to correlate
with higher number of citations~\cite{bornmann08}.
It is natural to define a relative indicator, the number of citations
per author. Is this normalization
the correct one that leads to a universal distribution,
for any number of authors?

Finally, from a more theoretical point of view, an interesting
goal for future work is to understand the origin
of the universality found and how its precise functional form comes
about. An attempt to investigate what mechanisms are relevant
for understanding citation distributions is in Ref.~\cite{vanraan01}.
Further activity in the same direction would be definitely interesting.

\section{Methods}
\label{methods}

Our empirical analysis is based on data from Thomson Scientific's
{\it Web of Science} (WOS, www.isiknowledge.com) database,
where the number of citations is counted as the total number of times
an article appears as a reference of a more recent published article.
Scientific journals are divided in 172 categories, from ``Acoustics'' to
``Zoology''. Within a single category a list of journals is provided.
We consider articles published in each of these journals
to be part of the category. Notice that the division in categories
is not mutually exclusive: for example {\it Physical Review D} belongs both to 
the ``Astronomy \& Astrophysics'' and to the ``Physics, particles \& fields''
categories.
For consistency, among all records contained in the database we consider
only those classified as ``article'' and ``letter'', thus
excluding reviews, editorials, comments and other published material
likely to have an uncommon citation pattern.
A list of the categories considered, with the relevant parameters that
characterize them, is reported in Table~\ref{table}.
The category "Multidisciplinary sciences" does not fit perfectly into
the universal picture found for other categories, because the distribution
of the number of citations is a convolution of the distributions
corresponding to the single disciplines represented in the journals. 
However, if one restricts only to the three most important multidisciplinary
journals ({\it Nature}, {\it Science}, {\it Proc. Natl. Acad. Sci. USA})
also this category fits very well into the global universal picture.

Our calculations neglect uncited articles; we have verified however that their 
inclusion just produces a small shift in $c_0$, which does not affect
the results of our analysis.
In the plots of the citation distributions, data have been grouped in bins of
exponentially growing size, so that they are equally spaced along a logarithmic
axis. For each bin, we count the number of articles with citation count
within the bin and  divide by the number of all potential values for the
citation count that fall in the bin (i.e. all integers). 
This holds as well for the distribution of the normalized citation
count $c_f$, as the latter is just determined by dividing the citation
count by the constant $c_0$, so it is a discrete variable just like the
original citation count. The resulting ratios obtained for each bin
are finally divided by the total number of articles considered,
so that the histograms are normalized to 1.





\begin{thebibliography}{}



\bibitem{hirsch05}
Hirsch JE (2005) An index to quantify an individual's scientific research output. {\it Proc Nat Acad Sci USA} 102:16569-16572.


\bibitem{egghe06}
Egghe L (2006) Theory and practise of the g-index. {\it Scientometrics} 69:131-152.


\bibitem{hirsch07}
Hirsch JE (2007) Does the h index have predictive power? {\it Proc. Nat. Acad. Sci. USA} 104:19193-19198.

\bibitem{evidence07}
Evidence Report (2007) {\scriptsize {\tt http://bookshop.universitiesuk.ac.uk/downloads/bibliometrics.pdf}}


\bibitem{kinney07}
Kinney AL (2007) National scientific facilities and their science impact on nonbiomedical research. {\it Proc Nat Acad Sci USA} 104: 17943-17947.

\bibitem{king04}
King DA (2004) The scientific impact of nations. {\it Nature} 430:311-316.


\bibitem{brooks86}
Brooks TA (1986) Evidence of complex citer motivations. {\it J Am Soc Inf Sci} 37:34-36.


\bibitem{egghe90}
Egghe L, Rousseau R (1990) in {\it Introduction to Informetrics: quantitative
methods in library, documentation and information science} (Elsevier, Amsterdam).


\bibitem{adler08}
Adler R, Ewing J, Taylor P (2008) Citation Statistics. {\it IMU Report}  
{\scriptsize {\tt http://www.mathunion.org/Publications/Report/CitationStatistics}}


\bibitem{bornmann08}
Bornmann L, Daniel H-D (2008) What do citation counts measure? A review of studies on citing behavior. {\it J Docum} 64:45-80.


\bibitem{althouse08}
Althouse BM, West JD, Bergstrom T, Bergstrom CT (2008) Differences in Impact Factor Across Fields and Over Time. {\it arxiv 0804.3116}.



\bibitem{garfield79}
Garfield E (1979) in {\it Citation Indexing. Its Theory and Applications in Science, Technology, and Humanities} (Wiley, New York).




\bibitem{schubert86}
Schubert A, Braun T (1986) Relative indicators and relational charts for comparative-assessment of publication output and citation impact. {\it Scientometrics} 9:281-291.


\bibitem{schubert96}
Schubert A, Braun T (1996) Cross-field normalization of scientometric indicators. {\it Scientometrics} 36:311-324.




\bibitem{vinkler96}
Vinkler P (1996) Model for quantitative selection of relative scientometric impact indicators. {\it Scientometrics} 36:223-236.




\bibitem{vinkler03}
Vinkler P (2003) Relations of relative scientometric indicators. {\it Scientometrics} 58:687-694.





\bibitem{iglesias07}
Iglesias JE, Pecharroman C (2007) Scaling the h-index for different scientific ISI fields. {\it Scientometrics} 73:303-320.






\bibitem{zitt05}
Zitt M, Ramanana-Rahary S, Bassecoulard E (2005) Relativity of citation performance and excellence measures: From cross-field to cross-scale effects of field-normalisation. {\it Scientometrics} 63:373-401.




\bibitem{redner98}
Redner S (1998) How popular is your paper? {\it Eur Phys J B} 4:131-134.





\bibitem{naranan71}
Naranan S (1971) Power law relations in science bibliography: a self-consistent interpretation. {\it J Docum} 27:83-97.




\bibitem{seglen92}
Seglen PO (1992) The skewness of science. {\it J Am Soc Inf Sci} 43:628-638.





\bibitem{fortunato07}
Fortunato S, Castellano C (2007) Scaling and Universality in Proportional Elections. {\it Phys Rev Lett} 99:138701.


\bibitem{plerou99}
Plerou V, Nunes Amaral LA, Gopikrishnan P, Meyer M, Stanley HE (1999) Similarities between the growth dynamics of university research and of competitive economic activities. {\it Nature} 400:433-437. 


\bibitem{lundberg07}
Lundberg J (2007) Lifting the crown—citation z-score. {\it J Informetrics} 1:145-154.




\bibitem{moed95}
Moed HF, Debruin RE, Vanleeuwen TN (1995) New bibliometric tools for the assessment of national research performance - database description, overview of indicators and first applications. {\it Scientometrics} 33:381-422.


\bibitem{vanraan04}
Van Raan AF (2004) Sleeping Beauties in Science. {\it Scientometrics} 59:461-466.


\bibitem{redner05}
Redner S (2005) Citation statistics from 110 years of Physical Review. {\it Phys Today} 58:49-54.

\bibitem{ball05}
Ball P (2005) Index aims for fair ranking of scientists {\it Nature} 436:900-900.

\bibitem{vanraan01}
Van Raan AF (2001) Competition amongst scientists for publication status: toward a model of scientific publication and citation distributions. {\it Scientometrics} 51:347-357.


\end{thebibliography}
\end{document}